    \documentclass[prd,aps,preprint,tightenlines,showpacs,nofootinbib,superscriptaddress]{revtex4-1}
\usepackage{mathrsfs}
\usepackage{amsfonts}
\usepackage{amsmath}
\usepackage{amssymb}
\usepackage{array}
\usepackage{verbatim}
\usepackage{bm}
\usepackage{epsfig}
\usepackage{graphicx,color}
\usepackage{relsize}
\usepackage{lineno}
\usepackage{float}
\usepackage{multirow}
\RequirePackage{xspace}

\def\lsim{\raise0.3ex\hbox{$<$\kern-0.75em\raise-1.1ex\hbox{$\sim$}}}

\def\gsim{\raise0.3ex\hbox{$>$\kern-0.75em\raise-1.1ex\hbox{$\sim$}}}

\newcommand{\be}{\begin{equation}}

\newcommand{\ee}{\end{equation}}

\def\beq{\begin{equation}}

\def\eeq{\end{equation}}

\def\beqa{\begin{eqnarray}}

\def\eeqa{\end{eqnarray}}

\newcommand{\ba}{\begin{eqnarray}}

\newcommand{\ea}{\end{eqnarray}}

\def\gappeq{\mathrel{\rlap {\raise.5ex\hbox{$>$}}

{\lower.5ex\hbox{$\sim$}}}}

\def\lappeq{\mathrel{\rlap{\raise.5ex\hbox{$<$}}

{\lower.5ex\hbox{$\sim$}}}}

\def\Toprel#1\over#2{\mathrel{\mathop{#2}\limits^{#1}}}

\begin{document}

\title{Particle production by $\gamma$-$\gamma$ interactions in future \\  electron-ion colliders}

\author{Carlos A. {\sc Bertulani}}
\email{carlos.bertulani@tamuc.edu}
\affiliation{Department of Physics and Astronomy, Texas A\&M University-Commerce, Commerce, TX 75429, USA. }

\author{Reinaldo {\sc Francener}}
\email{reinaldofrancener@gmail.com}
\affiliation{Instituto de Física Gleb Wataghin - Universidade Estadual de Campinas (UNICAMP), \\ 13083-859, Campinas, SP, Brazil. }

\author{Victor P. {\sc Gon\c{c}alves}}
\email{barros@ufpel.edu.br}
\affiliation{Institute of Physics and Mathematics, Federal University of Pelotas (UFPel), \\
  Postal Code 354,  96010-900, Pelotas, RS, Brazil}

\author{Juciene T. de {\sc Souza}}
\email{juciteixeiraprof@gmail.com}
\affiliation{Institute of Physics and Mathematics, Federal University of Pelotas (UFPel), \\
  Postal Code 354,  96010-900, Pelotas, RS, Brazil}

\begin{abstract}
The particle production in photon-photon ($\gamma \gamma$) interactions present in  electron-ion collisions is investigated. We present calculations for the  total cross sections and event rates related to the production of light mesons [$\eta, \eta^\prime, f_0$ and $f_2$], charmonium [$\eta_c$ and $\chi_c$] and charmoniumlike [$X(3915), X(3940), X(4140)$ and $X(6900)$] states, considering the EIC, EicC, LHeC and FCC-eh energies. 
Our predictions demonstrate that experimental studies of these processes are feasible and useful to
constrain the properties of light mesons and quarkonium states and shed some light on the configuration of the considered charmoniumlike
 states.
\end{abstract}



\maketitle

\vspace{1cm}

\section{Introduction}

\begin{figure}[b]
	\centering
\includegraphics[width=0.65\textwidth]{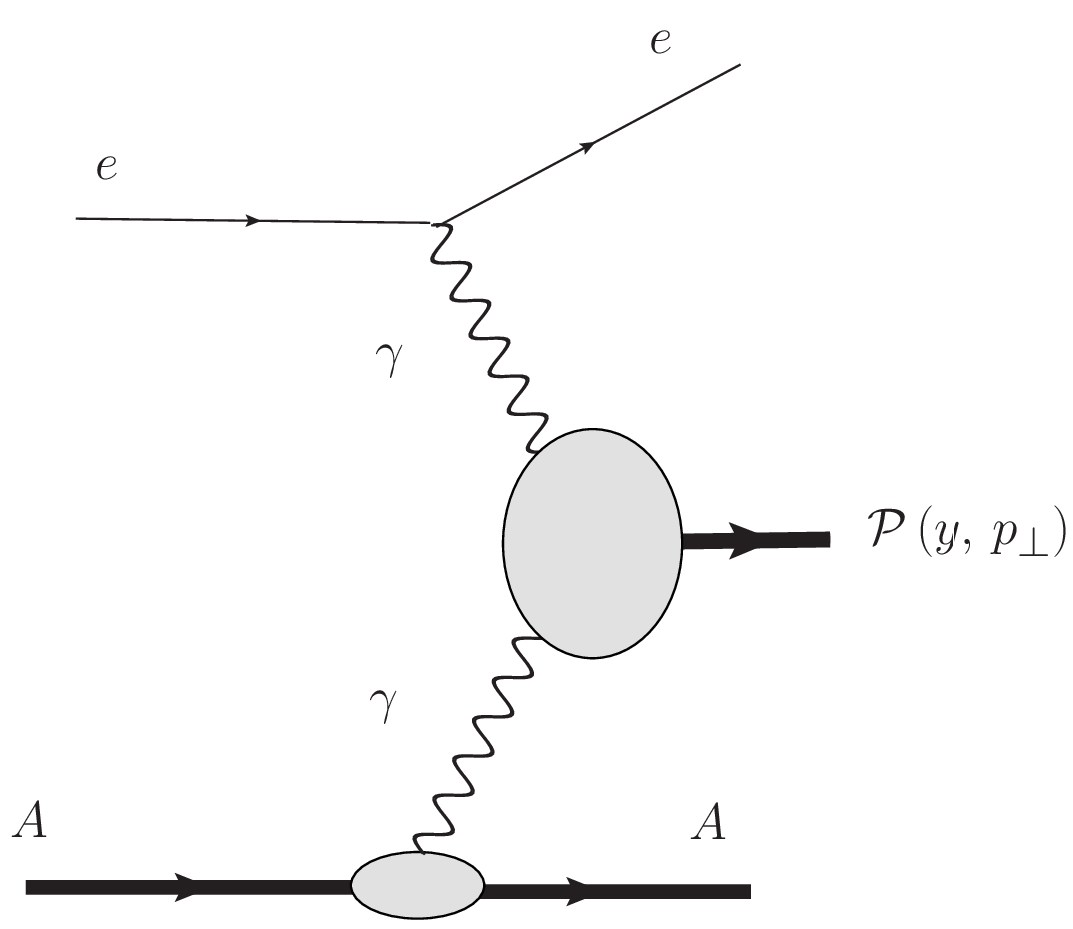} 
\caption{Particle production by $\gamma \gamma$ interactions in eA  collisions. }
\label{fig:diagram}
\end{figure}

Exotic hadrons cannot  be easily accommodated in unfilled  $q\bar{q}$ and $qqq$ states. This fact has been established over the last years (For  reviews see, e.g., Refs. \cite{Karliner:2017qhf,Olsen:2017bmm,Liu:2019zoy,Chen:2022asf}).
In particular,  several candidates have been observed at the LHC, through the analysis of the decays from the particles produced in $pp$ collisions.  However, in recent years, the possibility of  probing exotic hadrons in photon induced interactions  at the LHC has been proposed and developed \cite{Natale:1995np,Schramm:1999tt,Natale:2001zj,Bertulani:2009qj,Machado:2010qg,vicwer,vicmar,nosexotico,Klein:2019avl,Goncalves:2018hiw,Goncalves:2019vvo,Xie:2020ckr,Goncalves:2021ytq,Xie:2021sik,Esposito:2021ptx,Biloshytskyi:2022dmo,Dong:2023eou,Feng:2023ghc,Fariello:2023uvh}. Such results indicated that the study of particle production by photon-photon and photon-hadron interactions at the LHC are an important alternative to  prove (or disprove) the existence  of these states and to investigate their properties. Considering that these interactions will also occur in the future electron-ion colliders, proposed to be constructed in the USA (EIC/BNL) \cite{eic}, Switzerland (LHeC/CERN and FCC-eh/CERN) \cite{lhec,fcceh} and China (EicC) \cite{EicC},   our objective in this paper is to extend these published studies and derive predictions for total cross sections and expected number of events considering some particular final states. We will focus on the particle production by $\gamma \gamma$ interactions in electron-nucleus collisions, as represented in Fig. \ref{fig:diagram}, which has an associated cross section factorized as products of the equivalent flux of photons of the incident particles  and  the  photon-photon production cross section. As the photon flux is well known, this process is sensitive to the description of production of the corresponding $  \gamma \gamma \rightarrow {\cal{P}}$ process and, consequently, to the  particle wave function.
In our analysis, in addition to the exotic charmoniumlike states X(3915), X(3940), X(4140) and X(6900), we will also consider the $\chi_c$ and $\eta_c$ charmonium states, which are an important background for the searching of an Odderon\footnote{The odderon is a state predicted by Quantum Chromodynamics
(QCD), characterized by a C-odd parity. It determines the cross section difference between the crossed and direct  channel processes at very high energies (see, e.g., Ref. \cite{ewerz}). In perturbative QCD, it is a C-odd composite state of three reggeized gluons described within  the
Bartels-Kwiecinski-Praszalowicz (BKP) formalism \cite{bkp}, which can be probed in the exclusive photoproduction of pseudoscalar mesons (See, e.g., Refs. \cite{ckms,bbcv,vic_odderon,Dumitru:2019qec,Benic:2023ybl,Benic:2024pqe}).} in $eA$ collisions, as well as some examples of light states ($\eta, \eta^{\prime}, f_0$ and $f_2$), whose future measurement can be useful to improve its description.   

This paper is organized as follows. In the following section, we present a brief review of the particle production formalism due to $\gamma \gamma$ interactions in $eA$ collisions. In particular, we will discuss the photon fluxes generated by an electron and a nucleus, as well as the expression for the photon-photon production cross section. In section \ref{sec:results}, we will present our calculations for the total cross sections and number of events considering the center-of-mass energies and luminosities expected for the future electron-ion colliders (EIC, LHeC, FCC-eh and EicC). In addition, we will consider the decay of the considered states in two photons and will present the associated predictions derived assuming  kinematical limits of the rapidities and energies of photons. A comparison with the predictions for the light-by-light (LbL) process will also be discussed. In the last section \ref{sec:sum}, we summarize our main predictions and conclusions.

\section{Formalism}

As discussed in the introduction, in the equivalent photon approximation (EPA) \cite{epa}, the cross section  for particle production by $\gamma \gamma$ interactions in $eA$ collisions is factorized in terms of a flux of equivalent photons on the incident particles  multiplied by  photon-photon production cross sections, i.e.,
\begin{eqnarray}
\sigma \left[e A \rightarrow e \otimes {\cal{P}} \otimes  A;\sqrt{s} \right]   
&=& \int \mbox{d}\omega_e 
\mbox{d}\omega_A \,   f_{\gamma/e}(\omega_e) f_{\gamma/A}(\omega_A)  \, \hat{\sigma}\left[\gamma \gamma \rightarrow {\cal{P}} ; 
W_{\gamma \gamma} \right] ,
\label{Eq:cs_singlet}
\end{eqnarray}
where $\otimes$ represents a rapidity gap in the final state, $\sqrt{s}$ is the center-of-mass energy in the $eA$ collision and $f_{\gamma/i}$ is the distribution function  of photons generated by particle $i$ ($i = e, \,A$) with a photon energy  $\omega_i$. Here,   $\hat{\sigma}$ is the cross section  for particle production in a  $\gamma \gamma$ interaction with a  particular photon-photon center-of-mass energy $W_{\gamma \gamma}$.

In our analysis, we take the flux associated with the electron as given by \cite{epa}
\begin{eqnarray}
    f_{\gamma/e}(\omega_e) =    \frac{\alpha_{em}}{\pi \omega_e} \int \frac{\mbox{d} Q^2}{Q^2}   \left[\left(1 - \frac{\omega_e}{E_e}\right)\left(1 - \frac{Q^2_{min}}{Q^2}\right) + \frac{\omega_e^2}{2E_e^2}\right]\,,
\end{eqnarray}
where $\omega_e$ represents the energy of the photon generated by the electron with a bombarding energy $E_e$, and $Q^2$ represents its virtuality. Moreover, kinematics imply that the minimum momentum transfer is given by $Q_{min}^2 = m_e^2\omega_e^2/[E_e(E_e - \omega_e)]$, whereas the maximum momentum transfer is  $Q_{max}^2 = 4E_e(E_e - \omega_e)$, due to the maximum of the electron energy loss in the process. In this paper, we will focus on the interaction between two real photons,  assuming $Q_{max}^2 = 1.0$ GeV$^2$ for the maximum momentum of the photon generated by the electron. It is important to emphasize that the study of $\gamma^* \gamma$ interactions in $eA$ collisions allow us to constrain the description of the meson transition form factors, as demonstrated in Ref. \cite{Babiarz:2023cac}, but the analysis of this interesting case is beyond the scope of the current paper.  For the nucleus, we will assume that  the photon distribution is given by \cite{upc},
\begin{eqnarray}
    f_{\gamma/A}(\omega_A) = \frac{Z^{2}\alpha}{\pi^2}  \int \mbox{d}^2r \frac{1}{r^{2} v^{2}\omega_A}
 \cdot \left[
\int u^{2} J_{1}(u) 
F\left(
 \sqrt{\frac{\left( \frac{r\omega_A}{\gamma_L}\right)^{2} + u^{2}}{r^{2}}}
 \right )
\frac{1}{\left(\frac{r\omega_A}{\gamma_L}\right)^{2} + u^{2}} \mbox{d}u
\right]^{2},
\label{fluxo}
\end{eqnarray}
where $F(q)$ represents the  charge form factor of the nucleus,   $\gamma_L$ is the laboratory Lorentz factor and $v$ is the velocity of the nucleus. The photon spectrum generated by the nucleus will be estimated using a realistic form factor \cite{Klein:1999qj}, corresponding to a Fourier transform of a  Wood-Saxon charge density distribution of the nucleus \cite{Woods:1954zz}, determined by low energy elastic electron-nucleus experimental data.

The cross section  $\hat{\sigma}_{\gamma \gamma}$ for the photoproduction of $\cal{P}$ state in $\gamma \gamma$ interactions can be estimated, at the Born level, employing Low's formula \cite{Low}. Such formula expresses this cross section  as a function of the two-photon decay width $\Gamma_{\cal{P} \rightarrow \gamma \gamma}$, i.e.,
\begin{eqnarray}
 \hat{\sigma}_{\gamma \gamma \rightarrow {\cal{P}}}(\omega_{e},\omega_{A}) = 
8\pi^{2} (2J+1) \frac{\Gamma_{{\cal{P}} \rightarrow \gamma \gamma}}{M_{\cal{P}}} 
\delta(4\omega_{e}\omega_{A} - M_{\cal{P}}^{2}) \, ,
\label{Low_cs}
\end{eqnarray}
where $M_{\cal{P}}$ and $J$ are the mass and spin of the  produced particle, respectively.

One has that the final state will be characterized by the particle $\cal{P}$, which will be produced in a rapidity $y$ with transverse momentum $p_{\perp}$, two intact recoiled particles (electron and nucleus)  and the existence of two  rapidity gaps  which are empty  pseudo-rapidity regions separating the intact very forward moving particles from the produced particle ${\cal{P}}$ state. In the $eA$ center-of mass (c.m.) frame, the photon energies $\omega_i$ are expressed in terms of the rapidity
$y$ as follows 
\begin{eqnarray}
    \omega_e = \frac{M_{\cal{P}}}{2} e^{+y}   \,\,\,\,\,\, \mbox{and} \,\,\,\,\,\,  \omega_A = \frac{M_{\cal{P}}}{2} e^{-y} \,.
    \label{eq:omegas}
\end{eqnarray}
As we are considering that the ${\cal{P}}$ state is produced by the interaction between two real photons, the typical $p_{\perp}$ is expected to be very small.
Such characteristics of the final state can be used, in principle, to perform the experimental separation of the associated events.
Another motivation to study the particle production by $\gamma \gamma$ interactions in electron-ion colliders is that the production mechanism is sensitive to the  annihilation process, $ {\cal{P}} \rightarrow \gamma \gamma$, and therefore to the particle wave function. Hence, studies of particle production in photon-induced interactions provides a direct test of the modeling of the produced states. 

\begin{center}
	\begin{table}[t]
		\begin{tabular}{|c|c|c|}
			\hline
			\hline 
    \textbf{Particle} &  \textbf{Mass (MeV)}& \textbf{Decay width $\Gamma\gamma \gamma$ (keV)} \\ 
    \hline
     \hline
        $\eta (547)$ & 547.9 &0.515  \\ 
        \hline
        $\eta' (958)$ & 957.8  &4.28 \\ 
        \hline
             $f_{0}(980)$& 990.0 & 0.29  \\
             \hline
            $f_{2}(1270)$ & 1275.4 & 2.60  \\
            \hline
            \hline
             \hline
         $\eta_{c}(1S)$ & 2984.1 & 5.10  \\
         \hline
          $\chi_{c0}(1P)$& 3414.7 & 33.60  \\
          \hline
          $\chi_{c2}(1P)$& 3556.2 & 0.578  \\
          \hline
           $\eta_{c}(2S)$& 3637.7 & 1.3 \\
          \hline
           \hline
            \hline
           $X(3915)$& 3919.4 & 0.200 \\
         \hline
         $X(3940)$ & 3942.0 & 0.330  \\
         \hline
         $X(4140)$& 4146.0 & 0.630  \\
         \hline
         $X(6900)$& 6886.0 & 67.0  \\
  \hline
  \hline 			
		\end{tabular}
		\caption{Properties of the particles considered in our analysis. The decay widths for the $X$ states are the theoretical values presented in Refs. \cite{nosexotico,Biloshytskyi:2022dmo}. For the other states, the values are those presented in the PDG \cite{ParticleDataGroup:2022pth}.}
		\label{table:mesons}
	\end{table}
\end{center}

In our study, we estimate the total cross  sections to produce  the particles listed in Table \ref{table:mesons}, which can be classified in three different groups: (a) light mesons [$\eta, \eta^\prime, f_0$ and $f_2$]; (b) charmonium states [$\eta_c$ and $\chi_c$]; and (c) charmoniumlike states [$X(3915)$, $X(3940)$, $X(4140)$ and $X(6900)$].  
The selection of particles in the group (a)  is motivated by the fact that a future measurement of these particles can be useful to improve our understanding of its structure and about the QCD vacuum \cite{Gan:2020aco}. In the group (b), we have considered particles that can also be produced in a photon-odderon interaction \cite{vic_odderon,Dumitru:2019qec,Benic:2023ybl}. Therefore, a precise determination of the contribution associated with the  $\gamma \gamma$ production is fundamental to discriminate between these two channels and probe the existence of the odderon in $eA$ collisions. Finally, in the group (c), some examples of charmoniumlike exotic systems, whose description is still a theme of debate \cite{Chen:2022asf}, are considered in order to investigate if the probing of these states in the future $eA$ colliders is feasible. For the exotic $X$ mesons,  we will assume the theoretical values for $\Gamma_{\gamma \gamma}$ presented in Refs. \cite{nosexotico,Biloshytskyi:2022dmo}. In contrast, for the other states, we will consider the values present in the PDG \cite{ParticleDataGroup:2022pth}. 
Although our selection is arbitrary, it can be easily extended for other particles that decay into a two-photon system. 

\begin{center}
	\begin{table}[t]
		\begin{tabular}{|c|c|c|c|c|c|c|}
			\hline
			\hline 
                           & eAu (EicC)                                   & eAu (EIC)                                    & ePb (LHeC)                     & ePb (FCC-eh) \\
			\hline 	
			\hline 
$\eta(547)$      & 253.51                (12.67$\times 10^{9}$) & 2126.88               (2.13$\times 10^{11}$) & 7905.82 (7.91$\times 10^{9}$)  & 12299.90 (92.25$\times 10^{9}$) \tabularnewline
			\hline 
$\eta '(958)$    & 125.50                (6.28$\times 10^{9}$)  & 2126.32               (2.13$\times 10^{11}$) & 9403.02 (9.40$\times 10^{9}$)  & 15341.50 (0.11$\times 10^{11}$) \tabularnewline
			\hline
$f_0(980)$       & 7.25                  (0.36$\times 10^{9}$)  & 125.59                (12.56$\times 10^{9}$) & 568.22  (0.57$\times 10^{9}$)  & 925.65   (6.94$\times 10^{9}$)  \tabularnewline
			\hline 
$f_2(1270)$      & 76.90                 (3.85$\times 10^{9}$)  & 2096.30               (2.10$\times 10^{11}$) & 10418.05(10.40$\times 10^{9}$) & 17455.26 (1.31$\times 10^{11}$) \tabularnewline
    \hline 
    \hline
    \hline
$\eta_{c}(1S)$   & 17.53$\times 10^{-3}$ (0.88$\times 10^{6}$)  & 23.41                 (2.34$\times 10^{9}$)  & 194.22  (0.19$\times 10^{9}$)  & 356.98   (2.68$\times 10^{9}$)  \tabularnewline
			\hline 
$\chi_{c0}(1P)$  & 19.15$\times 10^{-3}$ (0.96$\times 10^{6}$)  & 85.20                 (8.52$\times 10^{9}$)  & 777.34  (0.78$\times 10^{9}$)  & 1466.98  (11.00$\times 10^{9}$) \tabularnewline
			\hline 
$\chi_{c2}(1P)$  & 0.68$\times 10^{-3}$  (34.00$\times 10^{3}$) & 6.16                  (0.62$\times 10^{9}$)  & 57.81   (57.81$\times 10^{6}$) & 109.62   (0.82$\times 10^{9}$)  \tabularnewline
			\hline 
$\eta_{c}(2S)$   & 0.25$\times 10^{-3}$  (12.55$\times 10^{3}$) & 2.44                  (0.24$\times 10^{9}$)  & 23.95   (23.95$\times 10^{6}$) & 45.29    (0.34$\times 10^{9}$)  \tabularnewline
    \hline
    \hline
    \hline 
$X(3915)$        & 1.40$\times 10^{-5}$  (0.70$\times 10^{3}$)  & 0.27                  (27.00$\times 10^{6}$) & 2.79    (2.79$\times 10^{6}$)  & 5.38     (5.38$\times 10^{6}$)  \tabularnewline
			\hline 
$X(3940)$        & 2.22$\times 10^{-5}$  (1.11$\times 10^{3}$)  & 0.44                  (44.00$\times 10^{6}$) & 4.51    (4.51$\times 10^{6}$)  & 8.70     (65.25$\times 10^{6}$) \tabularnewline
			\hline 
$X(4140)$        & 2.87$\times 10^{-5}$  (1.44$\times 10^{3}$)  & 0.65                  (65.00$\times 10^{6}$) & 7.12    (7.12$\times 10^{6}$)  & 13.85    (0.10$\times 10^{9}$)  \tabularnewline
			\hline 
$X(6900)$        & 2.28$\times 10^{-6}$  (114.15)               & 5.37                  (0.54$\times 10^{9}$)  & 111.67  (0.11$\times 10^{9}$)  & 240.70   (1.81$\times 10^{9}$)  \tabularnewline
			\hline 
			\hline 			
		\end{tabular}
		\caption{Total cross sections in nanobarns (event rates per year)  for particle production via $\gamma \gamma$ interactions in eA collisions.}
		\label{table:total_nocuts}
	\end{table}
\end{center}

\section{Results}
\label{sec:results}

Here we estimate the total cross  sections considering energy and target configurations planned for future electron-ion colliders at the BNL, CERN and in China. 
In a future electron-ion collider at BNL,  electron beams with energies up to 18 GeV are expected to collide with heavy ions with energies below 100 GeV \cite{eic}. The colliding beams will reach  luminosities in the $10^{33} - 10^{34}$ cm$^{-2}$s$^{-1}$ interval. In our calculations, we assume  electron and $Au$-ion energies:  $(E_e,\, E_{Au}) = (18,\, 100)$ GeV, as a typical example.  Moreover, we will also estimate the cross  sections for the EicC \cite{EicC} ($E_e = 3.5$ GeV, $E_{Au} = 10$ GeV and ${\cal{L}} = 10^{33}$ cm$^{-2}$s$^{-1})$,  for the  LHeC \cite{lhec} ($E_e = 50$ GeV, $E_{Pb} = 2760$ GeV and   ${\cal{L}} = 10^{32}$ cm$^{-2}$s$^{-1}$) and for the  FCC-eh \cite{fcceh} ($E_e = 60$ GeV, $E_{Pb} = 19500$ GeV and   ${\cal{L}} = 54 \times 10^{32}$ cm$^{-2}$s$^{-1}$). These values imply that $\sqrt{s}|_{FCC-eh}  > \sqrt{s}|_{LHeC} > \sqrt{s}|_{EIC} > \sqrt{s}|_{EicC}$. 
The corresponding results are presented in Table \ref{table:total_nocuts}.

One has that  the cross section for a given final state increases with energy, which is directly associated with the increasing of the number of photons available for the $\gamma \gamma$ interaction.  
For the light mesons, we predict cross  sections of the order of $\mu$b and $\approx 10^{11}\, (10^{19}\,)$ events per year at the EIC (LHeC). In contrast for the charmonium (charmoniumlike) states, our calculations show a reduction by a factor $10^2$ ($10^3$), except for $X(6900)$ production.
At the EicC,  the production of massive states is strongly suppressed due to the small phase space available.

\begin{figure}[t]
	\centering
\includegraphics[width=0.8\textwidth]{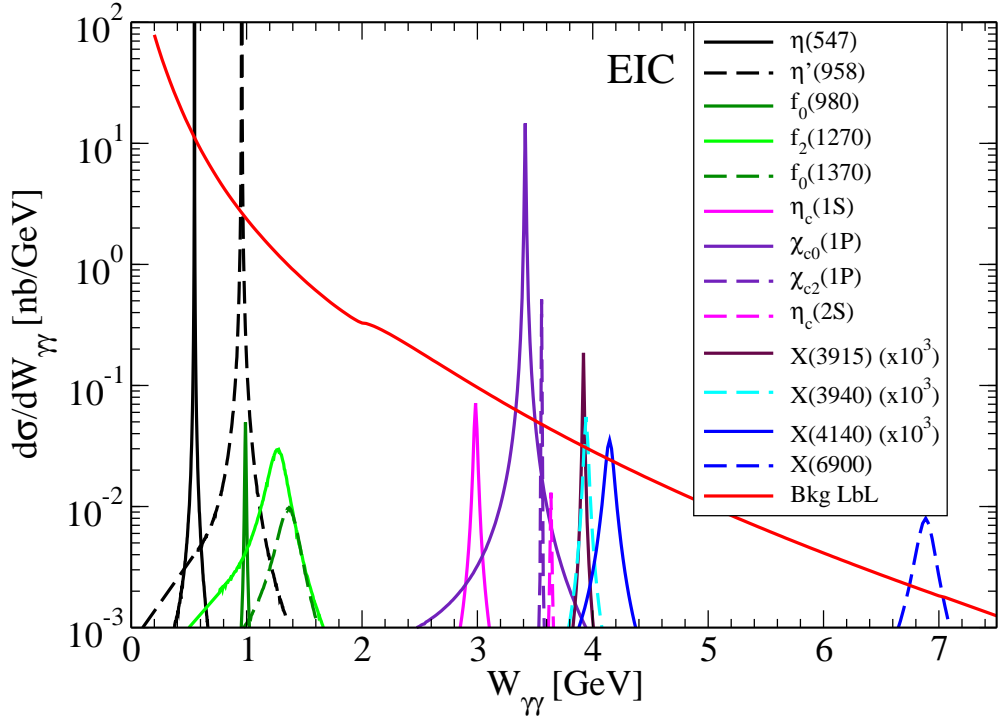} 
\caption{Predictions for the differential distribution, $d\sigma/dW_{\gamma \gamma}$, as a function of $W_{\gamma \gamma}$ for distinct particles decaying into a two-photon system.   For comparison, the differential distribution associated with the LbL processes is also presented. Results for the EIC energy.  }
\label{fig:spectrum}
\end{figure}

The results presented above motivate an analysis of the experimental separation of events. In what follows, we will consider that after being produced, the particles decay back in photons. Consequently,  the final state includes the electron, the ion, the two photons  and the occurrence of   two rapidity gaps. The invariant mass of the two photons will peak at $W_{\gamma \gamma} \approx M_{\cal{P}}$. Such behavior is observed in Fig. \ref{fig:spectrum}, where we present the differential distribution, $d\sigma/dW_{\gamma \gamma}$, as a function of $W_{\gamma \gamma}$ for distinct particles decaying into two photons, calculated for the EIC energy. For comparison, the differential distribution associated with the LbL processes is also presented. One has that in several cases the LbL contribution for the diphoton production is larger than that from the particle decay. In order to verify if it is possible to separate these two contributions,  we will study the consequences of kinematical cuts for the photon pseudorapidities and energies  in our calculations. The following constraints will be considered:
\begin{eqnarray}
    |\eta_1|, |\eta_2|  \le  3.5  \,\,\,\,\,\, \mbox{and} \,\,\,\,\,\,
    E_{\gamma_1}, E_{\gamma_2}  \ge  1.0 \,\, \mbox{GeV}   \,\,,
\end{eqnarray}
which are expected to hold at the EIC and are related with the acceptance region of the EIC calorimeters  \cite{eic}. The two cuts allow the detection of the photons  by the central detector and guarantee that they  are energetic enough to be reconstructed with the calorimeter. We will also apply similar cuts  in our predictions for the EicC, LHeC and FCC-eh.  The decay of the $\cal{P}$ states are simulated with the Pythia6 event generator \cite{Sjostrand:2006za}.
In Table \ref{table:withcuts} we show our obtained predictions for the total cross sections and events rates per year for the production of a two-photon system after the decay of a  $\cal{P}$ state produced in an $eA$ collision.
In comparison with the results presented in Table \ref{table:total_nocuts}, one observes that the cuts and the decay induce a suppression in the number of events by approximately three orders of magnitude. However, our results indicate that the number of events per year associated with the production of light mesons and charmonium states at the EIC, LHeC and FCC-eh, will be huge, which will allow, in principle, a future experimental analysis of these states. For the charmoniumlike states, the values assumed for $\Gamma_{\gamma \gamma}$ imply that these could be investigated in $eA$ collisions at the EIC and FCC-eh.

\begin{center}
	\begin{table}[t]
 \footnotesize     
		\begin{tabular}{|c|c|c|c|c|c|c|}
			\hline
			\hline 
                           & eAu (EicC)                                    & eAu (EIC)                                    & ePb (LHeC)                                   & ePb (FCC-eh) \\
			\hline 	
			\hline 
$\eta(547)$      & 5.74                  (0.29$\times 10^{9}$)   & 88.60                 (8.86$\times 10^{9}$)  & 319.60                (0.32$\times 10^{9}$)  & 457.29                (3.35$\times 10^{9}$)  \tabularnewline
			\hline 
$\eta '(958)$     & 0.24                  (12.00$\times 10^{6}$)  & 8.34                  (0.83$\times 10^{9}$)  & 34.55                 (34.55$\times 10^{6}$) & 51.51                 (0.39$\times 10^{9}$)  \tabularnewline
			\hline 
$f_0(980)$       & 2.44$\times 10^{-5}$  (1.22$\times 10^{3}$)   & 0.80$\times 10^{-3}$  (80.00$\times 10^{3}$) & 3.37$\times 10^{-3}$  (3.37$\times 10^{3}$)  & 5.27$\times 10^{-3}$  (39.53$\times 10^{3}$) \tabularnewline
			\hline 
$f_2(1270)$      & 19.80$\times 10^{-5}$ (9.80$\times 10^{3}$)   & 8.55$\times 10^{-3}$  (0.85$\times 10^{6}$)  & 39.05$\times 10^{-3}$  (39.05$\times 10^{3}$) & 59.35$\times 10^{-3}$ (0.45$\times 10^{6}$) \tabularnewline
    \hline 
    \hline
    \hline
$\eta_{c}(1S)$   & 2.38$\times 10^{-6}$  (0.12$\times 10^{3}$)   & 3.34$\times 10^{-3}$  (0.33$\times 10^{6}$)  & 24.81$\times 10^{-3}$ (24.81$\times 10^{3}$) & 40.00$\times 10^{-3}$ (0.30$\times 10^{6}$)  \tabularnewline
			\hline 
$\chi_{c0}(1P)$  & 57.33$\times 10^{-6}$ (2.87$\times 10^{3}$)   & 0.24                  (24.00$\times 10^{6}$) & 2.01                  (2.01$\times 10^{6}$)  & 3.29                  (24.68$\times 10^{6}$) \tabularnewline
			\hline 
$\chi_{c2}(1P)$  & 0.19$\times 10^{-6}$  (9.50)                  & 1.61$\times 10^{-3}$  (0.16$\times 10^{6}$)  & 13.79$\times 10^{-3}$ (13.79$\times 10^{3}$) & 22.76$\times 10^{-3}$ (0.17$\times 10^{6}$)  \tabularnewline
			\hline 
$\eta_{c}(2S)$   & 27.28$\times 10^{-9}$ (1.35)                  & 0.24$\times 10^{-3}$  (24.35$\times 10^{3}$) & 2.17$\times 10^{-3}$  (2.17$\times 10^{3}$)  & 3.60$\times 10^{-3}$  (26.97$\times 10^{3}$) \tabularnewline
    \hline
    \hline
    \hline 
$X(3915)$        & 2.15$\times 10^{-10}$ (10.75$\times 10^{-3}$) & 3.84$\times 10^{-6}$  (0.38$\times 10^{3}$)  & 36.31$\times 10^{-6}$ (36.31)                & 60.88$\times 10^{-6}$ (0.46$\times 10^{3}$)  \tabularnewline
			\hline 
$X(3940)$        & 1.87$\times 10^{-10}$ (9.35$\times 10^{-3}$)  & 3.40$\times 10^{-6}$  (0.34$\times 10^{3}$)  & 3.23$\times 10^{-5}$  (32.30)                & 5.42$\times 10^{-5}$  (0.41$\times 10^{3}$)  \tabularnewline
			\hline 
$X(4140)$        & 2.17$\times 10^{-10}$ (10.85$\times 10^{-3}$) & 4.59$\times 10^{-6}$  (0.46$\times 10^{3}$)  & 46.50$\times 10^{-6}$ (46.50)                & 78.82$\times 10^{-6}$ (0.59$\times 10^{3}$)  \tabularnewline
			\hline 
$X(6900)$        & 0.91$\times 10^{-9}$  (45.26$\times 10^{-3}$) & 2.11$\times 10^{-3}$  (0.21$\times 10^{6}$)  & 41.87$\times 10^{-3}$ (41.87$\times 10^{3}$) & 77.91$\times 10^{-3}$ (0.58$\times 10^{6}$)  \tabularnewline
			\hline 
			\hline 			
		\end{tabular}
		\caption{Cross sections in nanobarns (event rates per year) for the production of a system of two photons from the decay of a meson created by $\gamma \gamma$ interactions in $eA$ collisions. We consider kinematical cuts in pseudorapidity and energy of each photon after the decay.}
		\label{table:withcuts}
	\end{table}
\end{center}

\begin{center}
	\begin{table}[t]
 \footnotesize     
		\begin{tabular}{|c|c|c|c|c|}
			\hline
			\hline 
                     Central mass      & eAu (EicC)                                   & eAu (EIC)                                    & ePb (LHeC)                                   & ePb (FCC-eh) \\
			\hline 	
			\hline 	 	
$M_{\eta(547)}$      & 17.98$\times 10^{-3}$ (0.90$\times 10^{6}$)  & 0.29                  (29.21$\times 10^{6}$) & 1.01                  (1.01$\times 10^{6}$)  & 1.45                  (10.86$\times 10^{6}$) 
\tabularnewline
			\hline 
$M_{\eta^\prime(958)}$    & 3.88$\times 10^{-3}$  (0.19$\times 10^{6}$)  & 0.12                  (12.25$\times 10^{6}$) & 0.50                  (0.50$\times 10^{6}$)  & 0.75                  (5.63$\times 10^{6}$)
\tabularnewline
			\hline 
$f_0(980)$       & 3.54$\times 10^{-3}$  (0.18$\times 10^{6}$)  & 0.11                  (11.48$\times 10^{6}$) & 0.48                  (0.48$\times 10^{6}$)  & 0.71                  (5.35$\times 10^{6}$) 
\tabularnewline
			\hline  
$M_{f_2(1270)}$      & 1.57$\times 10^{-3}$  (78.55$\times 10^{3}$) & 73.24$\times 10^{-3}$ (7.32$\times 10^{6}$)  & 0.33                  (0.33$\times 10^{6}$)  & 0.51                  (3.79$\times 10^{6}$) 
\tabularnewline
    \hline 
    \hline
    \hline
$M_{\eta_c(1S)}$     & 10.14$\times 10^{-6}$ (507.21)               & 14.10$\times 10^{-3}$ (1.41$\times 10^{6}$)  & 0.10                  (0.10$\times 10^{6}$)  & 0.17                  (1.26$\times 10^{6}$)
\tabularnewline
			\hline 
$M_{\chi_{c0}(1P)}$  & 1.95$\times 10^{-6}$  (97.59)                & 9.32$\times 10^{-3}$  (0.93$\times 10^{6}$)  & 77.12$\times 10^{-3}$ (77.12$\times 10^{3}$) & 0.13                  (0.95$\times 10^{6}$) 
\tabularnewline
			\hline 
$M_{\chi_{c2}(1P)}$  & 1.15$\times 10^{-6}$  (57.97)                & 8.16$\times 10^{-3}$  (0.82$\times 10^{6}$)  & 70.61$\times 10^{-3}$ (70.61$\times 10^{3}$) & 0.12                  (0.87$\times 10^{6}$) 
\tabularnewline
			\hline 
$M_{\eta_{c}(2S)}$   & 0.87$\times 10^{-6}$  (43.63)                & 7.59$\times 10^{-3}$  (0.76$\times 10^{6}$)  & 66.94$\times 10^{-3}$ (66.94$\times 10^{3}$) & 0.11                  (0.83$\times 10^{6}$) 
\tabularnewline
    \hline
    \hline
    \hline
$M_{X(3915)}$        & 0.37$\times 10^{-6}$  (18.64)                & 5.90$\times 10^{-3}$  (0.59$\times 10^{6}$)  & 56.26$\times 10^{-3}$ (56.26$\times 10^{3}$) & 94.23$\times 10^{-3}$ (0.71$\times 10^{6}$) 
\tabularnewline
			\hline 
$M_{X(3940)}$        & 0.35$\times 10^{-6}$  (17.57)                & 5.79$\times 10^{-3}$  (0.58$\times 10^{6}$)  & 55.50$\times 10^{-3}$ (55.49$\times 10^{3}$) & 93.00$\times 10^{-3}$ (0.71$\times 10^{6}$)
\tabularnewline
			\hline 
$M_{X(4140)}$        & 0.21$\times 10^{-6}$  (10.67)                & 4.85$\times 10^{-3}$  (0.49$\times 10^{6}$)  & 49.14$\times 10^{-3}$ (49.14$\times 10^{3}$) & 82.93$\times 10^{-3}$ (0.62$\times 10^{6}$) 
\tabularnewline
			\hline 
$M_{X(6900)}$        & 0.25$\times 10^{-9}$  (12.28$\times 10^{-3}$)& 0.66$\times 10^{-3}$  (66.39$\times 10^{3}$) & 13.15$\times 10^{-3}$ (13.15$\times 10^{3}$) & 24.59$\times 10^{-3}$ (0.18$\times 10^{6}$)                 
\tabularnewline
			\hline 
			\hline 		
		\end{tabular}
		\caption{Cross-sections in nb (Event Rates per Year) for LbL scattering in eA collisions. Predictions derived considering cuts in the pseudorapidity and energy of each final photon, as well as in the invariant mass of the diphoton system, $W_{\gamma \gamma} = M_{\cal{P}} \pm 2.4\%M_{\cal{P}}$, with the central value presented in the first column. }
		\label{table:LbL}
	\end{table}
\end{center}

Finally, we include the contribution of the continuum due to LbL scattering and we present our results in Table \ref{table:LbL}, which has been estimated assuming the mentioned cuts on the energies of the final state photons and taking into account the invariant mass of the two-photon system, $W_{\gamma \gamma}$, within the range $M_{\cal{P}} \pm 2.4\%M_{\cal{P}}$ \cite{Francener:2024eep}. One observes that the LbL predictions are smaller (same order) than the results for the production of light mesons (charmonium states). In contrast, the LbL predictions are larger than those for the production of charmoniumlike states, except for the $X(6900)$ state. However, it is important to emphasize that the continuum background can be strongly decresed by measuring the LbL scattering in a resonance-free region, e.g., in an invariant mass range below and above of the peak determined by the $\cal{P}$ state, which enables a constraint in the magnitude of the peak. As a byproduct, the  contribution of LbL events could be largely reduced. 
As the number of LbL events  in electron-ion collisions is expected to be very large, such methodology will be feasible, in principle.  Our calculations therefore strongly motivate to proceed with a more detailed analysis with a Monte Carlo implementation of the mechanisms considered in our study and the potential emerging backgrounds. We plan to perform such an analysis in a forthcoming study.

\section{Summary}
\label{sec:sum}

In this paper we carried out an exploratory study of the particle production by $\gamma \gamma$ interactions in electron-ion collisions at the EicC, EIC,  LHeC and FCC-eh.  We predicted total cross sections and event rates per year. In addition, we considered the decay of the produced states in a diphoton system using kinematical cuts on the rapidities and energies of the photons. We have also demonstrated that a large number of events associated with light mesons and charmonium states is expected in the future colliders, which will allow us to improve our understanding about its structure and properties. In addition, the probing of charmoniumlike states will also be, in principle, feasible. In particular, our results point out that the EIC is a potential collider to produce exotic states. Our predictions  strongly motivate the implementation of dedicated Monte Carlo calculations in the description of $eA$ collisions and an extension of the study to  particle production in the interaction between a real and a virtual photon. Both subjects will be explored in forthcoming publications.

\begin{acknowledgments}
V. P. G. thanks the members of the Department of Physics and Astronomy of the Texas A \& M University-Commerce by the warm hospitality during the initial phase of this project and the IANN-QCD network by the financial support for the visit to that institution. 
R. F. acknowledges support from the Conselho Nacional de Desenvolvimento Cient\'{\i}fico e Tecnol\'ogico (CNPq, Brazil), Grant No. 161770/2022-3. V.P.G. and J. T. de Souza were partially supported by CNPq, CAPES (Finance Code 001), FAPERGS and INCT-FNA (Process No. 464898/2014-5). C.A.B acknowledges support by the U.S. DOE Grant DE-FG02-08ER41533 and the Helmholtz Research Academy Hesse for FAIR.

\end{acknowledgments}

\hspace{1.0cm}

\end{document}